\input{aipcheck}

\documentclass[
    ,final            % use final for the camera ready runs
  ]
  {aipproc}

\layoutstyle{6x9}

\begin{document}

\title{The nearby universe observed in far-infrared and in ultraviolet: an
analysis of the dust attenuation and the star formation activity}

\author{V\'eronique Buat \& the GALEX team}{
  address={Laboratoire d'Astrophysique de Marseille, France}
}

%\author{and the GALEX team}{
%  address={}
  %,altaddress={<author1 address>} % additional visiting address
%}

\begin{abstract}
 We discuss the dust attenuation and the star formation rates in the nearby
universe obtained from a comparison of far-infrared (IRAS) and ultraviolet
(GALEX) observations. The ratio of the dust to UV flux ratio is used to derive
the dust attenuation: this dust attenuation is found to increase with the
luminosity of the galaxies and from z=0 to z=1. The slope of the UV continuum
is found  to be a  very poor tracer of the dust attenuation in "normal"
galaxies.
Galaxies selected by their UV emission are found to be rather quiescent with a
recent star formation rate equal to only 25-30\% of the past averaged one.
Galaxies selected in FIR appear slightly more active in star formation.   
\end{abstract}
\maketitle

%%%%%%%%%%%%%%%%%%%%%%%%%%%%%%%%%%%%%%%%%%%%
%% MAINMATTER
%%%%%%%%%%%%%%%%%%%%%%%%%%%%%%%%%%%%%%%%%%%%

\section{Introduction}

The measure of the star formation rate (SFR) in galaxies is based on the
analysis of the emission from young stars which escapes the galaxies or is
absorbed and  re-emitted by the gas or the dust. 
It might be thought that the most direct tracer is the UV light emitted by
young
stars but the attenuation of this UV light by the dust absorption prevents
from any quantitative estimate of the SFR with the UV emission if no
correction for this attenuation is applied. To estimate this dust
attenuation has long been recognized as a crucial issue. Calzetti, Meurer and
collaborators showed that the UV continuum from $\sim$ 1200 \AA
to $\sim$ 2500\AA can be fitted by a power-law in starburst galaxies
($f_{\lambda}\propto \lambda^{\beta}$) (\cite{calz}); the slope $\beta$ of the
UV continuum was found to be  tightly related to the dust attenuation in
nearby
starburst galaxies  (e.g.\cite{meureretal}). Nevertheless, it appeared that the
method
was not valid for all types of galaxies: nearby spirals as well as Ultra
Luminous Infrared Galaxies do not follow the starburst relation \cite{bell}
\cite{goldaderetal}. \\
Another way to estimate the dust attenuation is to perform an energetic budget:
the UV light which does not
escape the galaxy   is absorbed by the interstellar dust and re-emitted in the
far-IR. Therefore both emissions originate from the same stellar populations
and their comparison is a powerful tracer of the dust attenuation (
\cite{buatxu} \cite{gordonetal}). They are also
closely related to the recent star formation rate over similar timescales
(\cite{buatxu} \cite{kennicutt}). \\
In this paper, we
will combine the UV and IR emissions and we will
combine the new GALEX data together with the existing far-IR data from IRAS to
discuss the dust  attenuation and star formation activity of galaxies

\section{The GALEX and IRAS data}

We have worked on 600 deg$^2$ observed by GALEX in NUV ($\lambda$ = 2310 \AA )
and FUV ($\lambda$ =  1530 \AA)  to build two samples of
galaxies. The first one, called the {\sl NUV selected sample} includes all the
galaxies brighter than m(NUV) = 16 mag (AB scale), among the 88 selected
galaxies (excluding ellipticals and active galaxies) only 3 are not detected
by IRAS at 60 $\mu m$. The second sample, called the {\sl FIR selected sample}
is based on the IRAS PSCz (\cite{saundersetal}): 118 galaxies
from this catalog lie within our GALEX fields, only 1 is not detected in NUV. 

\section{Dust attenuation in galaxies}
\subsection{Mean values of the dust attenuation}
For both samples we measure the dust attenuation using the dust to UV flux
ratio. This ratio is a quantitative measure of the dust attenuation at UV
wavelengths. It has been shown to be robust against variations of the dust
properties and of the star formation rate as long as stars are formed actively
in galaxies(e.g. \cite{buatxu} \cite{bua05} 
\cite{gordonetal}). The formulae used here
are  obtained for the  GALEX 
bandpasses (\cite{bua05}). 
\begin{equation}
A({\rm FUV}) = -0.0333y^3+0.3522y^2+1.1960y+0.4967
\end{equation}
where $y=\log(F_{\rm dust}/F_{\rm FUV})$.
and 
\begin{equation}
A({\rm NUV}) = -0.0495 y^3+0.4718 y^2+0.8998 y+ 0.2269
\end{equation}
where $y=\log(F_{\rm dust}/F_{\rm NUV})$.
The total (8-1000 $\mu m$) dust
emission is calculated from the fluxes at 60 and 100 $\mu m$ using the Dale et
al. (\cite{daleetal}) recipe. The NUV fluxes are calculated as $\nu\times
f_{\nu}$ and therefore $F_{\rm dust}/F_{\rm FUV}$ and 
$F_{\rm dust}/F_{\rm NUV}$ are unitless\\.
A moderate attenuation is found in the NUV selected sample with
$0.8^{-0.3}_{+0.4}$  mag in NUV and  and $1.1^{-0.4}_{+0.5}$ mag in FUV. As
expected the dust attenuation is higher in the FIR selected sample with
$2.1^{-0.8}_{+1.2}$ mag in NUV and $2.9^{-1.2}_{+1.2}$ mag in FUV.\\

\subsection{Dust attenuation  \& bolometric luminosity of galaxies}
Hereafter the dust and the NUV luminosities are added to give an estimate  of
the
bolometric luminosity of the galaxies in Fig 1. This luminosity is plotted
against the
dust attenuation in NUV for the NUV selected and the FIR selected samples. A
clear increase of the dust attemuation with the luminosity of the galaxies is
obtained although the trend is dispersed for the IR selected sample. 

\begin{figure}
\resizebox{35pc}{!}{\includegraphics[angle=-90]{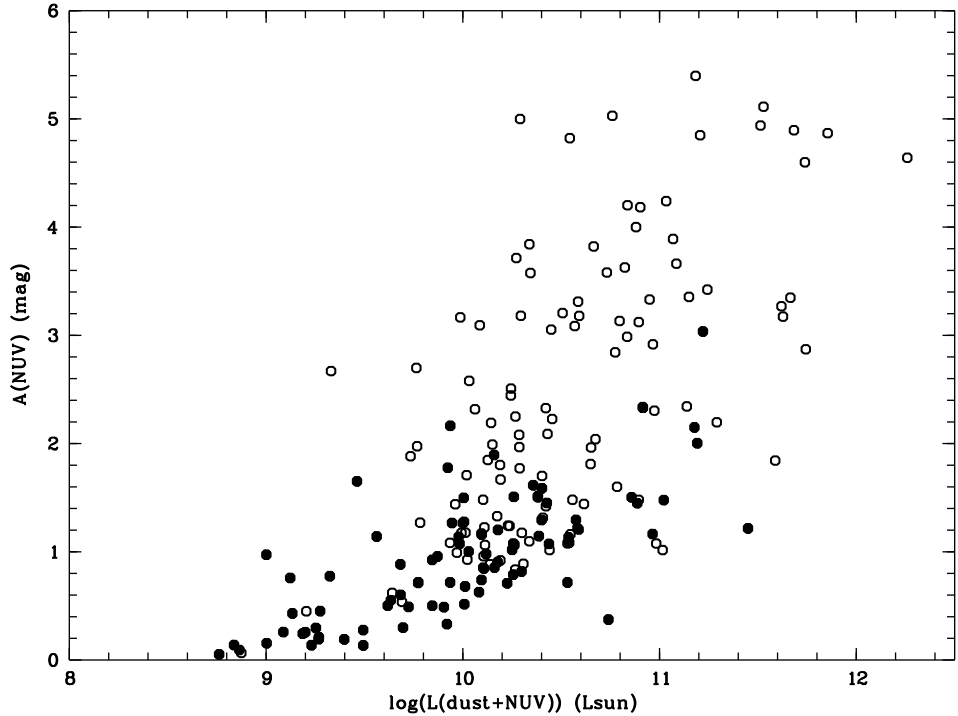}}
 \caption{Dust attenuation in the NUV band as a function of the total (dus
+NUV) luminosity of the galaxies. FIR selected galaxies are represented by
open circles, NUV selected galaxies by filled circles}
\end{figure}
\subsection{The slope of the UV continuum as a tracer of the dust attenuation}

The FUV-NUV color from the GALEX observations is directly linked to $\beta$
(\cite{kongetal}). We have plotted this color versus
F(dust)/F(FUV) for our two samples of galaxies in Fig. 2 and compared to the
predictions for starburst galaxies \cite{meureretal}.
Obviously $\beta$ is not a reliable tracer of the dust attenuation in our NUV
and FIR selected samples. Moreover, different behaviors are found within both
samples: whereas the FIR selected galaxies spread over a large area of the
diagram most of the NUV selected galaxies lie below the starburst relation.

\subsubsection{ The NUV selected sample}
Let us focus on the behavior of galaxies from the NUV selected sample. Kong et
al.  \cite{kongetal} have explored the effect of various star formation
histories
on the correlation found originaly for starburst galaxies. They showed that
assuming a decrease of the global star formation activity lead to redder
FUV-NUV colors for a similar dust attenuation as compared to a constant star
formation rate or a current star burst. In
Fig 2 we reproduce the results of their model obtained for a ratio of the
present to the past averaged SFR (the so called b parameter) equal to 0.25.
The model is consistent with  our data: the NUV selection does not seem to 
favour galaxies very active in star formation at least in the nearby universe. 
\begin{figure}
\resizebox{35pc}{!}{\includegraphics[angle=-90]{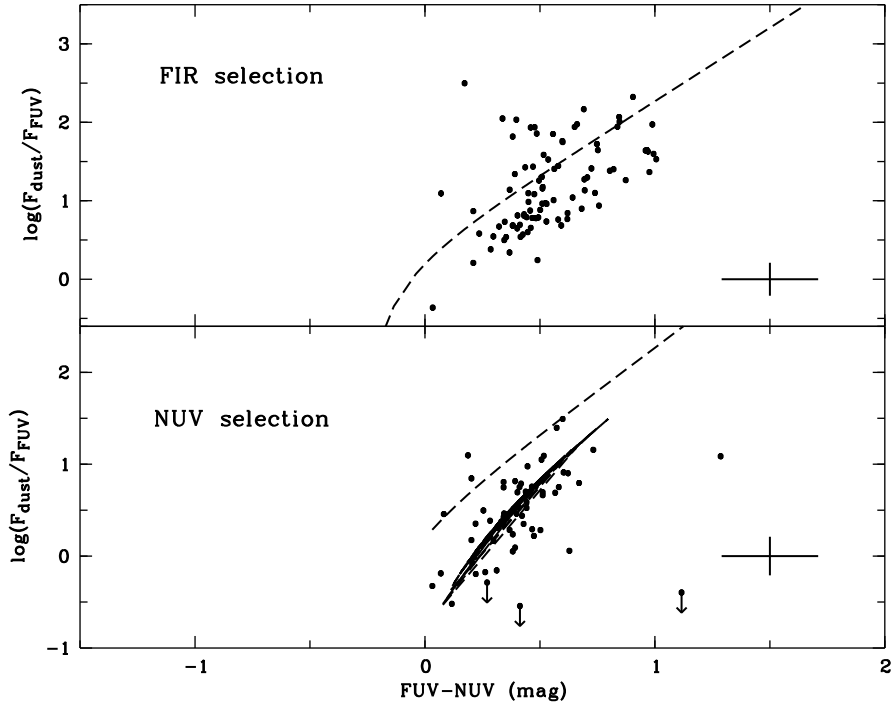}}
 \caption{ log($F(dust)/F(FUV)$) against the FUV-NUV color for NUV
and FIR
selected samples. The dashed line is the mean relation expected for starburst
galaxies, the  solid line is the locus of Kong et al. models
with b=0.25 (see text).
}
\end{figure}
\subsubsection{The effect of the dust properties}
Two factors are expected to have a large effect on the FUV-NUV color: the star
formation history and the dust properties. Kong et al \cite{kongetal} have
explored the first factor (see above) but a variation of the dust attenuation
curve has
also to be explored. We have used the Charlot and Fall \cite{charlot}
formalism and assumed a power law variation of the optical depth of the dust: 
$\tau(\lambda)\propto \lambda^{-n}$ allowing n to vary from 2 to 0.5 (whereas
Charlot and Fall fixed n to 0.7 from an analysis of starburst galaxies). The
NUV bandpass of GALEX is well centered on the Galactic bump at 2175 \AA,
therefore if such a bump exists in external galaxies it may have a strong
effect on  the
FUV-NUV color. Therefore we also simulate the bump by adding a Gaussian curve
centered on 2175 \AA (amplitude variable and $\sigma$ =200 \AA) to the power
law variation of the optical
depth   and simulating the bump. In Fig 3 are presented the results of this
model for a
continous star formation rate (using the PEGASE synthesis model \cite{fioc}).
The solid line is
the locus of starburst galaxies and the dashed one the prediction of Kong et
al for b=0.25.  Whereas the only way to obtain redder FUV-NUV colors with a low
dust attenuation seems to assume a quiescent current star formation,  a
variable dust attenuation curve alone (with a constant star formation rate)
 can  also explain  the behavior of many galaxies in Fig. 2 \\
Therefore a complete interpretation of the data will need to consider both 
temporal variations of the star formation history and spectral variation of 
the dust attenuation  among galaxies (Burgarella et al. in preparation).

\begin{figure}
\resizebox{25pc}{!}{\includegraphics[angle=-90]{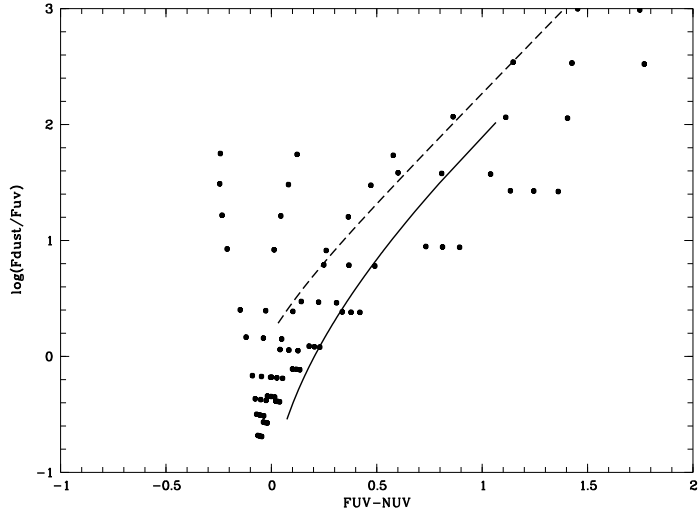}}
 \caption{Model predictions assuming a constant star formation rate and
different dust attenuation curves (see text for details)}
\end{figure}
\subsection{ The mean dust attenuation from z$\sim$0 to z$\sim$1}
A comparison of the FIR and UV luminosity densities from IRAS
(\cite{saundersetal2}) and GALEX (\cite{wyder}) leads to
a mean dust attenuation at z$\sim$0 of 1.1 mag in NUV and 1.5 mag in FUV: the
nearby universe is not very obscured \cite{bua05}.
With the recent results of GALEX we can also
perform the  analysis from z=0 to $z\sim 1$.The dust
emission is derived from  Chary \& Elbaz \cite{chary}: we translate the results
of their
model for the evolution of the obscured star formation rate in dust luminosity
density using the formulae given in their papers. 
The rest-frame FUV luminosity density as a function of z is taken from 
\citep{schimi}. The dust to FUV luminosity density ratios thus calculated
are translated in dust attenuation using the relation (1). The results are
gathered in Tab. 1. A clear increase of the
mean dust attenuation with the redshift is observed: the dust luminosity
density increases more than the UV one with the redshift 
leading to an increase of the global dust attenuation by $\sim$ 1 mag from z=0
to z>0.5. 
\begin{table}
\begin{tabular}{lrrrr}
\hline
   \tablehead{1}{r}{b}{redshift}&
  \tablehead{1}{r}{b}{$\rho(dust)$\\$L\odot Mpc^{-3}$}&
   \tablehead{1}{r}{b}{$\rho(FUV)$\\$L\odot Mpc^{-3}$}&
   \tablehead{1}{r}{b}{$\rho(dust)/\rho(FUV)$}  &
   \tablehead{1}{r}{b}{A(FUV)\\mag}\\
\hline
0.06& $9 10^7$ & $1.8 10^7$ & 5    & 1.5\\
0.3& $24 10^7$ & $3.8 10^7$ & 6.3 & 1.7\\
0.5 &  $65 10^7$& $4.2 10^7$  & 15.7 & 2.4\\
0.7 & $160 10^7$ & $7.7 10^7$& 20.8  & 2.6\\
1 & $204 10^7$ & $6.9 10^7$& 29.6  & 2.9\\
\hline
\end{tabular}
\caption{Evolution of the mean dust attenuation, the luminosity densities are
taken from \cite{schimi} in FUV and 
\cite{chary} for the dust emission.}
\label{tab:a}
\end{table}

\section{Recent star formation rate \&  star formation activity}

\subsection{The measure of the recent star formation rate}
UV and total dust  emissions can be calibrated in a recent star formation rate
assuming a star formation history over $\sim 10^8$ years  and an
initial mass
function  \cite{buatxu} \cite{kennicutt}. When using
the dust luminosity one must add an additional hypothesis about the absorption 
of the stellar light by the dust. The classical hypothesis (also made in this
work) is that all the stellar light from the young stars is absorbed by the
dust \cite{kennicutt}. In this paper we also assume  a Salpeter
IMF from 1 to 100 M$\odot$. Using Starburst99 synthesis models we
obtain:
\begin{equation}
log (L_{NUV}) (L\odot) = 9.73+log(SFR) (M\odot yr^{-1})
\end{equation}
 and
\begin{equation}
log (L_{dust}) (L\odot) = 10.168+log(SFR) (M\odot yr^{-1})
\end{equation}
\begin{figure}
\resizebox{35pc}{!}{\includegraphics[angle=-90]{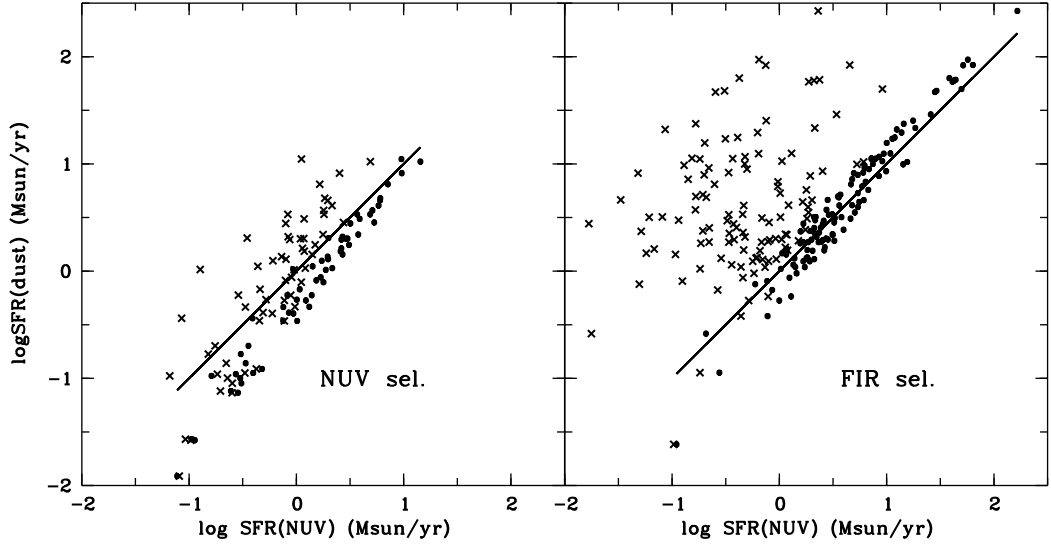}}
 \caption{SFR  deduced from the dust luminosity versus the SFR deduced from
the NUV luminosity directly observed (crosses) and corrected for dust
attenuation (points) for the NUV selected sample (left panel) and the FIR
selected one (right panel). The  lines correspond to equal quantities on
both axes}
\end{figure}
In Fig 4 are plotted the SFR estimated from the NUV luminosity against the SFR
from the dust luminosity. In both samples, the {\it observed} NUV luminosity
strongly under-estimates the SFR the effect being worse for the FIR selected
sample. When the UV fluxes of the FIR selected sample are corrected for dust
attenuation the
agreement is very good (as expected) between both estimates of the SFR  since
we are dominated in each case by the dust
emission. Conversely in the NUV selected sample the SFR estimated from the
dust luminosity alone is found to underestimate systematically the SFR as
compared to the NUV
corrected luminosity. The discrepancy increases towards the low SFRs to reach
a factor 3 for SFR of $\sim$ 0.3 M$\odot~ yr^{-1}$. Therefore, using the dust
emission alone to measure the total SFR in all galaxies can be misleading, the
best way being to combine UV and IR emissions to estimate reliable SFRs
\cite{iglesias}

 \subsection{Star formation activity}

\begin{figure}
\resizebox{30pc}{!}{\includegraphics[angle=-90]{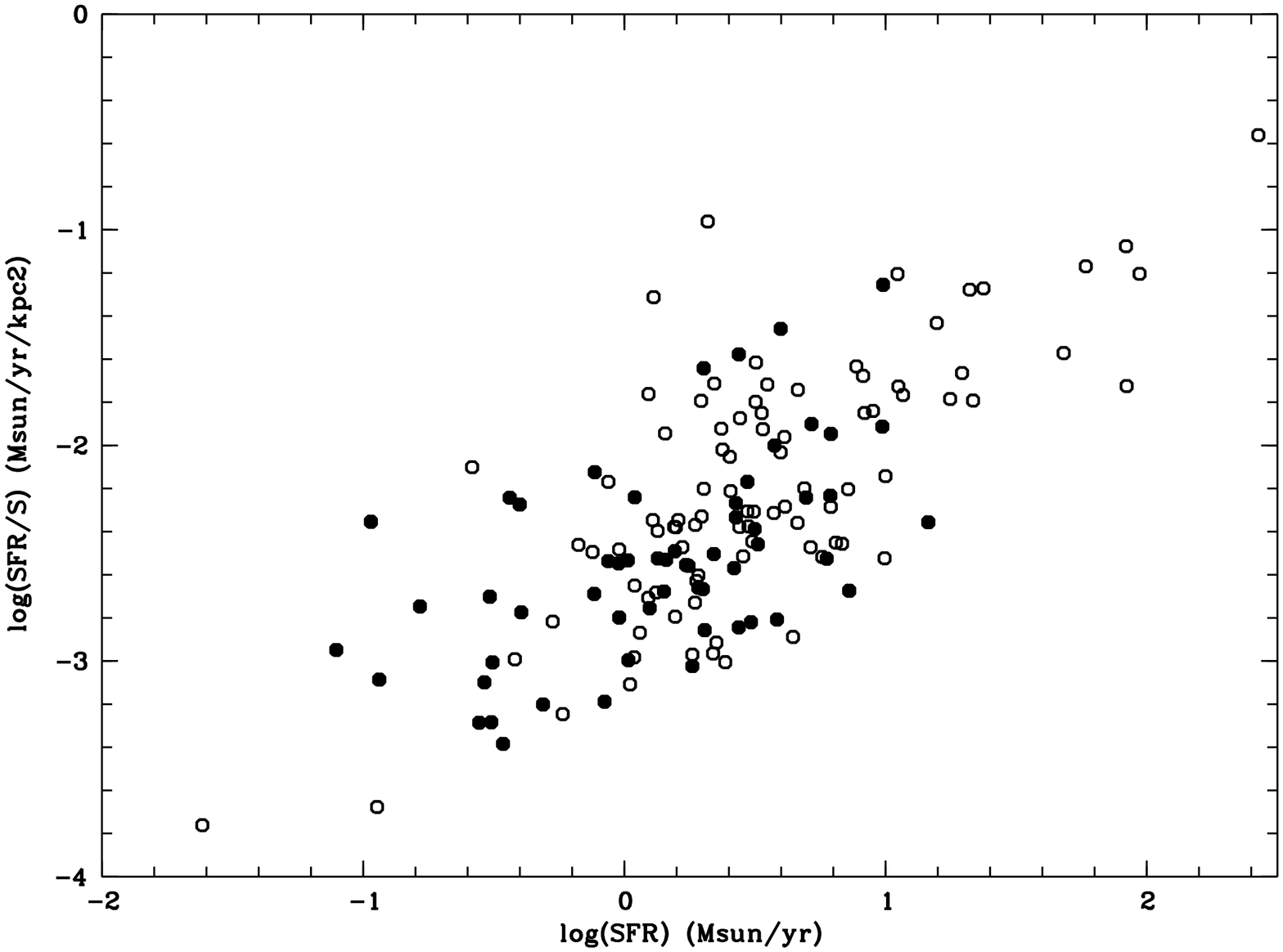}}
\caption{The star formation rate per unit area plotted against the total   
SFR for NUV selected galaxies (filled circles) and FIR selected galaxies 
(open circles)}
\end{figure}

We  can compare the strength of the star formation in the NUV and FIR selected
samples by plotting the SFR per unit area in both samples as a function of the
total star formation rate. For the NUV selected sample the SFR is obtained from
the NUV fluxes corrected for dust attenuation and using the formula (3); for
the FIR selected sample we use the total dust emission and formula (4). The
results are shown in Fig 5. The FIR selected
galaxies appear more active than the NUV selected ones both in terms of total
SFR and surface density of star formation. No strong starbursts is selected in
either sample: starbursts are characterized by a SFR per unit area larger than
0.1 $M\odot yr^{-1}$ (Kennicutt 2004, in Starbursts-From 30 Doradus to Lyman
Break Galaxies) and only two galaxies selected
in FIR exhibit such a value.\\
We can also calculate the b parameter (the ratio of the present SFR to the past 
averaged one) by using the H magnitude from the 2MASS survey. Using the
calibration from \cite{boselli} we obtain a mean b parameter of $0.3
\pm 0.4$ for the NUV selected sample and $0.4\pm 0.5$ for the FIR selected
sample (Fig. 6). Once again the FIR selected galaxies seem to be slightly more
active than the NUV selected ones. The b parameters found for the NUV selected
sample are consistent with the value of the model of Kong et al.
(\cite{kongetal}) used to fit 
our data in Fig 2.

\begin{figure}
\resizebox{30pc}{!}{\includegraphics[angle=-90]{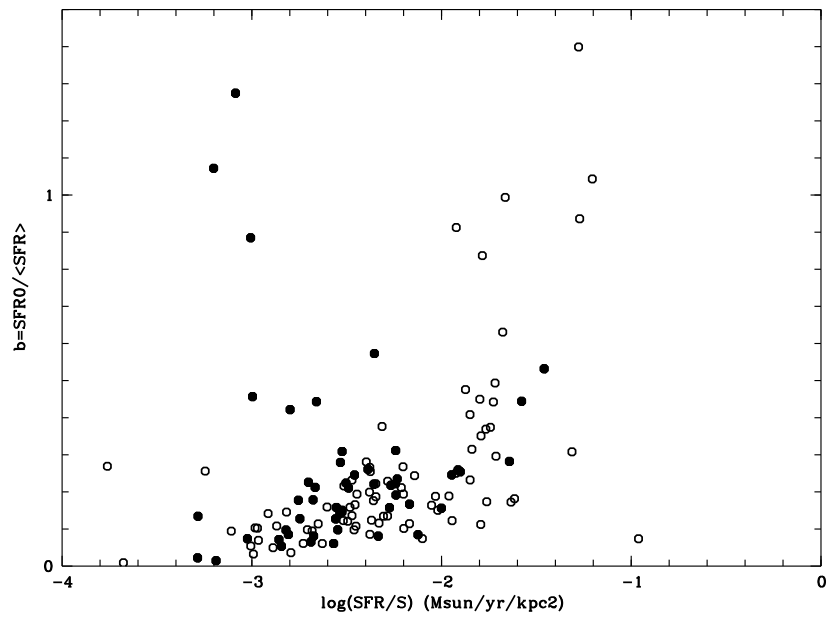}}
 \caption{The ratio of the present to past average SFR as a function of the SFR
per unit area for NUV selected galaxies (filled circles) and FIR selected
galaxies (open circles)}
\end{figure}

\section{conclusions}

 FIR and UV emissions are complementary to obtain reliable estimates of
the dust attenuation and of the star formation rate in galaxies. Using the FIR
or the UV alone can lead to errors: on one hand the star formation rate of
 galaxies selected in UV is under-estimated by a
factor which can reach  $\sim$3 using the FIR alone, on the other hand for the
vast majority of the
galaxies the UV must be corrected for dust attenuation before any quantitative
use. Therefore the best estimate of the star formation rate is probably to add
the observed  FIR and UV contributions.\\
A FIR or a NUV selection in the nearby universe does not select
starbursts, FIR selected galaxies are found slightly more active in SFR than
the NUV selected ones. NUV selected galaxies  have a mean ratio of the present
SFR to the past averaged one of 0.2-0.3\\

The dust attenuation measured by comparing the dust and UV emissions is found
to increase with the total luminosity of the galaxies. The interpretation of
the F(dust)/F(NUV) vs FUV-NUV diagramm  (i.e. the dust attenuation vs the
slope of
their UV continuum) is difficult and depends on the dust properties and the
star formation history in a complex way. \\
The mean dust attenuation appears to increase from z=0 to z$\sim$1 by
more than 1 mag.

%%%%%%%%%%%%%%%%%%%%%%%%%%%%%%%%%%%%%%%%%%%%%%%%
%% BACKMATTER
%%%%%%%%%%%%%%%%%%%%%%%%%%%%%%%%%%%%%%%%%%%%%%%%

\begin{theacknowledgments}
I thank Cristina Popescu and Richard Tuffs to have invited me to present our
most recent results from the GALEX observations. 
GALEX (Galaxy Evolution Explorer) is a NASA Small Explorer, launched in April
2003.
We gratefully acknowledge NASA's support for construction, operation,
and science analysis for the GALEX mission,
developed in cooperation with the Centre National d'Etudes Spatiales
of France and the Korean Ministry of 
Science and Technology. 
\end{theacknowledgments}

%% If the bibliography is produced without BibTeX comment out the

\bibliographystyle{aipproc}   % if natbib is available


\IfFileExists{\jobname.bbl}{}
 {\typeout{}
  \typeout{******************************************}
  \typeout{** Please run "bibtex \jobname" to optain}
  \typeout{** the bibliography and then re-run LaTeX}
  \typeout{** twice to fix the references!}
  \typeout{******************************************}
  \typeout{}
 }

\end{document}